\documentclass[a4paper]{article}
\usepackage[a4paper,top=3cm,bottom=2cm,left=3cm,right=3cm,marginparwidth=1.75cm]{geometry}
\usepackage{amsfonts}
\usepackage{bm}
\usepackage{extarrows}
\usepackage{amssymb}
\usepackage{color}
\usepackage[all]{xy}
\usepackage{graphicx}
\usepackage{braket}
\usepackage{amsmath}
\usepackage{appendix}
\usepackage{graphicx}
\usepackage[colorinlistoftodos]{todonotes}
\usepackage{amsthm}
\usepackage{mathrsfs}
\usepackage{amssymb}
\usepackage{accents}
\usepackage{extarrows}
\usepackage{makecell}
\usepackage{authblk}
\usepackage{url}
\usepackage{hyperref}
\usepackage{cite}
\usepackage{eufrak}
\hypersetup{colorlinks,
            citecolor=black,
            linkcolor=black,
            urlcolor=green,
            pdftex}

\usepackage[english]{babel}
\usepackage[utf8x]{inputenc}
\usepackage[T1]{fontenc}

\usepackage[normalem]{ulem}

\title{Polytopes in all dimensional loop quantum gravity}

\author[1,2]{Gaoping Long \footnote{201731140005@mail.bnu.edu.cn}}
\author[1,3]{Yongge Ma   \footnote{mayg@bnu.edu.cn}\thanks{corresponding author}}
\affil[1]{School of Physics and Technology, Xinjiang University, Urumqi 830046, China}
\affil[2]{Department of Physics, South China University of Technology, Guangzhou 510641, China}
\affil[3]{Department of Physics, Beijing Normal University, Beijing 100875, China}
\date{}
\begin{document}

\maketitle
\begin{abstract}
The Lasserre's reconstruction algorithm is extended to the D-polytopes with the construction of their shape space. Thus, the areas of d-skeletons $(1\leq d\leq D)$ can be expressed as  functions of the areas and normal bi-vectors of the (D-1)-faces of D-polytopes. As weak solutions of the simplicity constraints in all dimensional loop quantum gravity, the simple coherent intertwiners are employed to describe semiclassical D-polytopes. New general geometric operators based on D-polytopes are proposed by using the Lasserre's reconstruction algorithm and the coherent intertwiners. Such kind of geometric operators have expected semiclassical property by the definition. The consistent semiclassical limit with respect to the semiclassical D-polytopes can be obtained for the usual D-volume operator in all dimensional loop quantum gravity by fixing its undetermined regularization factor case by case.
\end{abstract}

\section{Introduction}
The kinematic structure of (1+3)-dimensional loop quantum gravity (LQG) \cite{Ashtekar2012Background}\cite{Han2005FUNDAMENTAL}\cite{thiemann2007modern}\cite{rovelli2007quantum} indicates a picture of quantum Riemann geometry, wherein the Hilbert space of this theory is given by the completion of the space of cylindrical functions on certain quantum configuration space. A cylindrical function $f_\gamma$ is based on a given graph $\gamma$ embedded in the 3-dimensional spatial hypersurface. The basic conceptions of the quantum Riemann geometry come from the 2-areas and 3-volumes of the dual polyhedra of the graph $\gamma$, which can be given by the spin representations labelling the edges and the intertwiners labelling the vertices of $\gamma$. In fact, for an $F$-valent vertex which is dual to a polyhedron with $F$ 2-faces, one could obtain a gauge-invariant intertwiner space $\mathcal{H}_F$ by quantizing certain classical phase space $\mathcal{S}_F$ \cite{baez1999quantum}\cite{barbieri1998quantum}\cite{bianchi2011polyhedra}, as introduced by Kapovich and Millson in \cite{kapovich1996symplectic}. The states of coherent intertwiners in $\mathcal{H}_F$ are labelled with the points in $\mathcal{S}_F$ and comprise an overcomplete basis in the kinematical Hilbert space of LQG based on the $F$-valent vertex. Moreover, the coherent intertwiner states can give the semiclassical 3-spatial geometry of polyhedra in large spin limit \cite{bianchi2011polyhedra}\cite{livine2007new}\cite{conrady2009quantum}.
The idea that the coherent intertwiners can be regarded as the semiclassical states of certain spatial geometry in large quantum number limit was also extended to all dimensional LQG \cite{bodendorfer2013newi}\cite{bodendorfer2013newiii}. It provides a way to deal with the anomalous quantum simplicity constraints in all dimensional LQG with certain geometrical meaning. In this way, the non-commutative quantum vertex simplicity constraints \cite{bodendorfer2013implementation} were imposed weakly \cite{Gaoping2019coherentintertwiner}, so that some kinds of weak solutions were constructed as well-behaved semiclassical states. These weak solutions are composed of $SO(D+1)$ coherent states \cite{perelomov2012generalized} and comprise the so-called simple coherent intertwiner space in all dimensional LQG.

A natural question is whether the simple coherent intertwiner space can also be obtain by quantizing the shape space of D-polytopes directly, just as the special case in (1+3)-dimensional theory. The answer is negative. In fact, the construction of Kapovich and Millson's phase space can be extended to higher dimensional cases but with some different characters as introduced in \cite{Gaoping2019coherentintertwiner}. In this description, in addition to the closure condition there exist extra simplicity constraints. These constraints lead to the fact that the shape space of D-polytopes is no longer a phase space. This may be taken as a new perspective to understand the anomaly of quantum vertex simplicity constraints. Hence the relation between the simple coherent intertwiners and D-polytopes in all dimensional LQG becomes worth studying. This is the main issue which will be studied in this paper. We will show that each gauge-invariant simple coherent intertwiner can be labelled by the geometric information of a D-polytope \cite{toth2017handbook}\cite{alexandrov2005convex}, similar to the (1+3)-dimensional case. In particular, thanks to the Minkowski Theorems \cite{alexandrov2004minkowski}\cite{minkowski1897allgemeine} and the properties of $SO(D+1)$ coherent states \cite{perelomov2012generalized}, we can show that each gauge-invariant simple coherent intertwiner is a quantum state describing a semiclassical polytope $\mathcal{P}$, and the product of the out-oriented normal bi-vector $V^{IJ}_\imath$ and the area $A_\imath$ of $\imath_{\textrm{th}}$ (D-1)-face of $\mathcal{P}$ can be promoted as an operator $X_\imath^{IJ}$ in the intertwiner space.

With the realization of the above quantum correspondence, how to represent the quantum geometric information of a simple coherent intertwiner describing a semiclassical polytope is the next valuable issue. In classical theory, the Minkowski Theorems ensure that all of the geometric information of a D-polytope is contained in the normal vectors $V^{IJ}_\imath$ and the areas $A_\imath$ of its (D-1)-faces, and the Lasserres's reconstruction algorithm \cite{lasserre1983analytical} provides a concrete way to express kinds of geometrical quantities of a D-polytope by $V^{IJ}_\imath$ and $A_\imath$. Also, notice that the product $A_\imath V^{IJ}_\imath$ is used to coordinatize the shape space of the D-polytopes as well as to label the simple coherent intertwiners. Since the geometrical quantities of the D-polytopes can be expressed as functions on their shape space, it is possible to define certain geometrical operators of the D-polytopes by the coherent intertwiners which are regarded as the coherent states of quantum D-polytopes. In fact, two kinds of general spatial geometrical operators in all dimensional LQG have been constructed in \cite{Gaoping2019geometricoperators}. One of them was constructed with the flux operators by the so-called internal regularization procedure. Some undetermined factors would be introduced in the expressions of this kind of operators. One way to fix the undetermined factors is to employ the semiclassical consistency check, which also requires to relate the geometrical quantities of a D-polytope to $V^{IJ}_\imath$ and $A_\imath $.
In this paper, in the light of the Minkowski theorems, we will extend the Lasserres's reconstruction algorithm to the D-polytopes adapted to the all dimensional LQG. Thus, the quantum theory of D-polytopes with the simple coherent intertwiners as coherent states can be obtained. Then, certain geometric operators will be defined by using the coherent states and their basic properties will be discussed. Besides, the undetermined regularization factor in the usual D-volume operator defined in \cite{bodendorfer2013newiii} will be fixed for two special cases of D-polytopes.

This paper is organized as follows. In section 2, we will introduce the space of shape of D-polytopes, which is constructed based on the normal bi-vector convention and the Minkowski theorems. Some properties of this shape space will be discussed. In section 3, we will show how to extend the Lasserres's reconstruction algorithm to arbitrary dimensions and give all the geometric information of a D-polytope by the areas and normals of its (D-1)-faces. Especially, the volume function on the shape space of two kinds of special D-polytope---D-simplex and parallel D-polytope---will be discussed in details. We will turn to quantum theory in section 4, in which the geometric quantization of the shape space of D-polytopes will be considered by neglecting the simplicity constraints. Then the quantum simplicity constraints will be imposed weakly on coherent intertwiners to get the simple coherent intertwiner space with minimal quantum uncertainty. Also, we will introduce the classical-quantum corresponding relation in this quantization scheme, which interprets the simple coherent intertwiners as states describing semiclassical D-polytopes. This corresponding relation provides a foundation to define new geometric operators in section 5, in which the new volume operator will be compared with the usual one defined in all dimensional LQG. The desired semiclassical behaviour of the usual volume operator with respect to the semicalssical D-polytopes in large $N$ limit will be obtained by fixing its undetermined regularization factor case by case. Finally, we will finish with a summary and discussion in section 6.
\section{The shape space of polytopes}
\subsection{Convex polytopes and Minkowski theorems}
A D-dimensional convex polytope is a convex hull of a finite set of points in D-dimensional Euclidean space $\mathbb{R}^D$. It can be represented as the intersection of finitely many half-spaces as \cite{alexandrov2005convex}
\begin{equation}
\mathcal{P}=\{x\in\mathbb{R}^D|V_\imath\cdot x\leq h_\imath,\imath=1,...,m\},
\end{equation}
where $V_\imath$ are arbitrary vectors, and $h_\imath$ are real numbers.  The above abstract description is non-unique and redundant, as the minimal set of half-spaces needed to describe a D-polytope corresponds to taking their number $m$ equal to the number of (D-1)-faces $F$ of the D-polytope. In this paper we are interested in the description of a convex D-polytope with $F$ faces in terms of variables that have an immediate geometric interpretation: the areas of the (D-1)-faces of the D-polytope and the unit normals to the (D-1)-hyperplanes that support these (D-1)-faces. Let us consider a set of unit vectors $V_\imath\in \mathbb{R}^D$ and  a set of positive real numbers $A_\imath$ such that they satisfy the $closure$ condition
\begin{equation}
 C\equiv\sum_{\imath=1}^FA_\imath V_\imath=0.
\end{equation}
 A convex polytope with $F$ (D-1)-faces having areas $A_\imath$ and normals $V_\imath$ can be obtained in the following way. For each vector $V_\imath$, one considers the (D-1)-hyperplane in $\mathbb{R}^D$ orthogonal it and translate it to a distance $h_\imath$ from the origin of $\mathbb{R}^D$. The intersection of the half-spaces bounded by these (D-1)-hyperplanes define the D-polytope, $V_\imath\cdot x\leq h_\imath$. Then, one can adjust the heights $h_\imath$ so that the (D-1)-hyperfaces have the expected areas $A_\imath$.

Remarkably, a convex D-polytope with such (D-1)-faces' areas and normals always exists. Moreover, it is unique, up to the global rotations and translations. This result is established by the following theorems due to H. Minkowski \cite{alexandrov2004minkowski}\cite{minkowski1897allgemeine}:

(i) \textbf{Theorem} (H. Minkowski uniqueness theorem).  Let $D\geq2$ and let two convex D-polytopes in $\mathbb{R}^D$ be such that, for every (D-1)-dimensional face of each of the D-polytope, the parallel face of the other D-polytope exists and has the same (D-1)-dimensional volume. Then the two D-polytopes are congruent and parallel to each other.

 (ii) \textbf{Theorem} (H. Minkowski existence theorem). Let $D\geq2$ and let $V_1,...,V_m$ be unit vectors in $\mathbb{R}^D$ that do not lie in a closed half-space bounded by a (D-1)-hyperplane passing through the origin. Let $A_1,...,A_m$ be positive real numbers such that $\sum_{\imath=1}^mA_\imath V_\imath=0$. Denote by $\vec{A}$ and $\vec{V}$ the sets $\{A_1,...,A_m\}$ and $\{V_1,...,V_m\}$ respectively. Then there exists a convex D-polytope $\mathcal{P}(\vec{A},\vec{V})$  in $\mathbb{R}^D$ such that its (D-1)-faces' number is $F=m$, the vectors $V_1,...,V_m$ (and only they) are the unit outward normal vectors to the (D-1)-dimensional faces of $\mathcal{P}(\vec{A},\vec{V})$, and the (D-1)-dimensional areas of the corresponding (D-1)-faces are equal to $A_1,...,A_m$ respectively.

The above discussions can be regarded as a direct generalization of the similar result for polyhedra in 3-dimensional Euclidean space, but the following discussions for general D-polytopes are different from 3-polyhedra.
 \subsection{Shape space of D-polytopes}
 The unit normal vector of a (D-1)-face in $\mathbb{R}^D$ is unique up to sign, and the configuration space of unit vectors in $\mathbb{R}^D$ is a (D-1)-dimensional unit sphere. In general,  symplectic structure does not exist for an arbitrarily dimensional sphere. Hence we can not construct the phase space of general D-polytopes based on the (D-1)-sphere. In order to give a general formulation, let us consider the D-polytopes in (D+1)-dimensional Euclidean space $\mathbb{R}^{(D+1)}$. Given an unit vector $n^I\in \mathbb{R}^{(D+1)}$, a subspace $\mathbb{R}_{n^I}^D\subset \mathbb{R}^{(D+1)}$ which satisfies $x^I n_I=0$, $\forall x^I\in\mathbb{R}_n^D$ can be fixed, wherein the index $I$ represents the abstract index to label the vector in $\mathbb{R}^{(D+1)}$. Following the result of last subsection, for two sets $\vec{A}$ and $\vec{V}^I$ satisfying $V^I_\imath n_I=0$, $\forall \imath=1,...,m$ and $\sum_{\imath=1}^mA_\imath V_\imath^I=0$, we have an unique D-polytope $\mathcal{P}(\vec{A},\vec{V}^I)$ in $\mathbb{R}_{n^I}^D$ up to translation. Notice that a (D-1)-face in $\mathbb{R}^{(D+1)}$ has an unique (up to sign) unit normal bi-vector $V_\imath^{[IJ]}$ and it takes the formulation $V_\imath^{[IJ]}=\sqrt{2}n^{[I}V_\imath^{J]}$ for the $\imath_{\textrm{th}}$ (D-1)-face of $\mathcal{P}(\vec{A},\vec{V}^I)$. Thus we can reformulate H. Minkowski theorem of D-polytopes in $\mathbb{R}^{(D+1)}$ as:

  \textbf{Theorem} (H. Minkowski theorem). Let $D\geq2$ and $n^I,V^I_1,...,V^I_m$ be unit vectors in $\mathbb{R}^{(D+1)}$ satisfying $n_IV^I_\imath=0$, $\forall \imath$, and $V^I_1,...,V^I_m$ span a D-dimensional vector space. Let $A_1,...,A_m$ be positive real numbers such that $\sum_{\imath=1}^mA_\imath V_\imath=0$ or equivalently $\sum_{\imath=1}^mA_\imath V^{IJ}_\imath=\sqrt{2}\sum_{\imath=1}^mA_\imath n^{[I}V_\imath^{J]}=0$. Denote by $\vec{A}$ and $\vec{V}^{IJ}$ the sets $\{A_1,...,A_m\}$ and $\{V^{IJ}_1=\sqrt{2}n^{[I}V_1^{J]},...,V^{IJ}_m=\sqrt{2}n^{[I}V_m^{J]}\}$ respectively. Then there exists an unique (up to translation) convex D-polytope $\mathcal{P}(\vec{A},\vec{V}^{IJ})$ in $\mathbb{R}^{(D+1)}$ with $F=m$ (D-1)-faces such that the bi-vectors $V^{IJ}_1,...,V^{IJ}_m$ (and only they) are the unit outward normal bi-vectors of the (D-1)-dimensional faces of $\mathcal{P}(\vec{A},\vec{V}^{IJ})$ and the (D-1)-dimensional areas of the corresponding (D-1)-faces are equal to $A_1,...,A_m$ respectively.

  In this formulation the areas and unit normal bi-vectors are used to label (D-1)-faces of D-polytopes in $\mathbb{R}^{(D+1)}$.  Let us consider the space of general bi-vectors $AV^{IJ}$ in $\mathbb{R}^{(D+1)}$ with norm $A$, which is given by $(Q_{D-1},\Omega_{A^2/2})$, where $Q_{D-1}:=SO(D+1)/(SO(2)\times SO(D-1))$ is a compact Kahler manifold and $\Omega_{A^2/2}:=A\Omega$ with $\Omega$ being the corresponding Kahler form on $Q_{D-1}$ \cite{mladenov1985geometric}. This space with the natural Poisson structure given by $\Omega_{A^2/2}$ can be regarded as the phase space of direction of a (D-1)-face with (D-1)-area $A$ in $\mathbb{R}^{D+1}$. Based on this result and the reformulated Minkowski theorem, we can give the space of shapes of D-polytopes with fixed (D-1)-areas $(A_1,A_2,...,A_F)$ of $F$ (D-1)-faces as \cite{Gaoping2019coherentintertwiner}\cite{Bodendorfer:2013sja}
\begin{eqnarray}\label{shapespace}
&&\mathfrak{P}^{s.}_{\vec{A}}:=\{(A_1V^{IJ}_1,A_2V^{IJ}_2,...,A_FV^{IJ}_F)\in Q_{D-1}(A_1)\times Q_{D-1}(A_2) \times\\\nonumber&&...\times Q_{D-1}(A_F)
| \sum_{\imath=1}^FA_\imath V^{IJ}_\imath=0,\quad V_\imath^{[IJ}V_\jmath^{KL]}=0\}/SO(D+1),
\end{eqnarray}
where $Q_{D-1}(A_\imath):=(Q_{D-1},\Omega_{A^2_\imath/2})$ is the phase space of $A_\imath V_\imath^{IJ}$, $V_\imath^{IJ}$ is the unit normal bi-vector of $\imath_{\textrm{th}}$ (D-1)-face, and
 \begin{equation}\label{clou}
 \sum_{\imath=1}^FA_\imath V^{IJ}_\imath=0
 \end{equation}
 is the closure condition,
 \begin{equation}\label{simpli}
   V_\imath^{[IJ}V_\jmath^{KL]}=0
 \end{equation}
is the so-called simplicity constraint which ensures that $V_\imath^{IJ}$ takes the formulation $V_\imath^{IJ}=\sqrt{2}n^{[I}V_\imath^{J]}$ such that the above reformulated Minkowski theorem is applicable. However this shape space is not a phase space and hence cannot be geometrically quantized. To overcome this problem, we first neglect the simplicity constraints and get another larger space
\begin{eqnarray}\label{Pphase}
&&\mathfrak{P}_{\vec{A}}:=\{(A_1V^{IJ}_1,A_2V^{IJ}_2,...,A_FV^{IJ}_F)\in Q_{D-1}(A_1)\times Q_{D-1}(A_2) \times\\\nonumber&&...\times Q_{D-1}(A_F)
| \sum_{\imath=1}^FA_\imath V^{IJ}_\imath=0,\}/SO(D+1),
\end{eqnarray}
which is indeed a phase space. The dimension of $\mathfrak{P}_{\vec{A}}$ reads $F(\frac{D(D+1)}{2}-1-\frac{(D-2)(D-1)}{2})-D(D+1)=2F(D-1)-D(D+1)$, and its Poisson structure is reduced from $\Omega_{A_\imath^2/2}$ on each of the compact Kahler manifold $Q_{D-1}$.
 \section{Reconstruction procedure}
So far we have introduced the conclusion that a D-polytope in $\mathbb{R}^{D+1}$ can be totally given by its  (D-1)-faces' areas and normal bi-vectors. We now describe how the D-polytope can be explicitly reconstructed from the (D-1)-areas and normals. The reconstruction will enable us to evaluate completely its geometry, including its D-volume and the areas of its skeletons. This can be done in two steps. In the first step, the algorithm due to Lasserre \cite{lasserre1983analytical} is extended to algebraically compute the areas and normal vectors of all the skeletons of the D-polytope as defined by $h_\imath$ and $V_\imath^I$. In the second step one solves a set of equations to obtain the values of the heights $h_\imath$ for the given areas of (D-1)-faces.
\subsection{Lasserre's reconstruction algorithm}
The Lasserre's procedure begins with a D-polytope $\mathcal{P}(\vec{h},\vec{V}^{IJ})$ defined by
\begin{equation}\label{def}
\mathcal{P}(\vec{h},\vec{V}^{IJ})=\mathcal{P}(\vec{h},\vec{V}):=\{x\in\mathbb{R}^D|V_\imath\cdot x\leq h_\imath, \imath=1,...,F\}
\end{equation}
with $F$ faces and $V_\imath^{IJ}=\sqrt{2}n^{[I}V_\imath^{J]}$, where $\vec{V}:=\{V^I_1,...,V_F^I\}$.
 Consider the $\imath_{\textrm{th}}$ (D-1)-face. The points $x\in\mathbb{R}^D_{n^I}$ on it satisfy
 \begin{eqnarray}\label{Vh}
  && V_\imath\cdot x= h_\imath, \\\nonumber
   && V_\jmath\cdot x\leq h_\jmath,\quad \forall\jmath\neq\imath.
 \end{eqnarray}
 We consider the generic case of $V_\imath\cdot V_\jmath\neq\pm1$, $\forall \imath, \jmath$ (the special configurations can be obtained as limit cases). It is convenient to introduce the coordinates $y_\imath$ adapted to the (D-1)-face such that
 \begin{equation}
 V_\imath\cdot y_\imath=0,\quad y_\imath=x-(x\cdot V_\imath)V_\imath.
 \end{equation}
 Using Eq. (\ref{Vh}) we have $x=h_\imath V_\imath+y_\imath$, and
 \begin{equation}
   y_\imath\cdot V_\jmath\leq r_{\imath\jmath},\quad \forall\jmath\neq\imath,
 \end{equation}
 wherein $r_{\imath\jmath}\equiv h_\jmath-(V_\imath\cdot V_\jmath)h_\imath$. Hence, the  $\imath_{\textrm{th}}$ (D-1)-face can be characterized either in terms of the $x$ or the $y_\imath$ coordinates by
 \begin{equation}
 \left\{
             \begin{array}{lr}
             x\cdot V_\imath=h_\imath&  \\
             x\cdot V_\jmath\leq h_\jmath, & \forall\jmath\neq\imath
             \end{array}
\right. \quad \rightarrow\quad \left\{
             \begin{array}{lr}
            y_\imath\cdot V_\imath=0&  \\
             y_\imath\cdot V_\jmath\leq r_{\imath\jmath}(h,V), & \forall\jmath\neq\imath.
             \end{array}
\right.
 \end{equation}
 Notice that $h_{\imath\jmath}:=r_{\imath\jmath}/\sqrt{1-(V_\imath\cdot V_\jmath)^2}$ is the distance of the (D-2)-face (skeleton) $[\imath\jmath]$ from the projection of the origin on the  $\imath_{\textrm{th}}$ (D-1)-face, where $[\imath\jmath]$ is the intersection set of $\imath_{\textrm{th}}$ and $\jmath_{\textrm{th}}$ (D-1)-faces. Note also that $[\imath\jmath]$ may not exist for certain $\imath$ and $\jmath$. Thus the conditions $y_\imath\cdot V_\jmath\leq r_{\imath\jmath}(h,V)$ are redundant if the $\jmath_{\textrm{th}}$ (D-1)-face does not intersect with $\imath_{\textrm{th}}$ (D-1)-face at a (D-2)-face. In this case the conditions will be excluded in the following discussion acquiescently. Besides, the normal vector $V_{\imath\jmath}$ of $[\imath\jmath]$ in the (D-1)-dimensional Euclidean space $\mathbb{R}_\imath^{D-1}$ in which the $\imath_{\textrm{th}}$ (D-1)-face lies is given by
 \begin{equation}
 V_{\imath\jmath}=\frac{V_\jmath-(V_\imath\cdot V_\jmath)V_\imath}{\sqrt{1-(V_\imath\cdot V_\jmath)^2}}.
 \end{equation}
Then the $\imath_{\textrm{th}}$ (D-1)-face as a (D-1)-polytope in $\mathbb{R}_\imath^{D-1}$ can be given by
\begin{equation}
 \mathcal{P}_{\imath}(\vec{h}_{\imath},\vec{V}_{\imath}):=\{y_\jmath\in\mathbb{R}_\jmath^{D-1}|V_{\imath\jmath}\cdot y_\jmath\leq h_{\imath\jmath}, \jmath=1,...,F_\imath\},
\end{equation}
where $\vec{h}_{\imath}:=(h_{\imath1},...,h_{\imath F_\imath})$, $\vec{V}_{\imath}:=(V_{\imath1},...,V_{\imath F_\imath})$, and $F_\imath$ is the number of the (D-2)-faces as the (D-2)-skeletons of the $\imath_{\textrm{th}}$ (D-1)-face. The above procedure of algorithm can be denoted by
\begin{equation}
\mathbf{L}_{\imath}^{D}:\quad \mathcal{P}(\vec{h},\vec{V})\Rightarrow\mathcal{P}_{\imath}(\vec{h}_{\imath},\vec{V}_{\imath}).
\end{equation}
Similarly, we can continue this procedure for (D-1)-polytope $\mathcal{P}_{\imath}(\vec{h}_{\imath},\vec{V}_{\imath})$ and get
\begin{equation}
\mathbf{L}_{\jmath}^{(D-1)}:\quad \mathcal{P}_{\imath}(\vec{h}_{\imath},\vec{V}_{\imath})\Rightarrow\mathcal{P}_{\imath\jmath} (\vec{h}_{\imath\jmath},\vec{V}_{\imath\jmath}),
\end{equation}
where $\mathcal{P}_{\imath\jmath} (\vec{h}_{\imath\jmath},\vec{V}_{\imath\jmath})$ is the $\jmath_{\textrm{th}}$ (D-2)-face of $\mathcal{P}_{\imath}(\vec{h}_{\imath},\vec{V}_{\imath})$, $\vec{V}_{\imath\jmath}=(...,V^I_{\imath\jmath\jmath'},...)$ is the set of the unit outward normal vectors of the (D-3)-skeletons of $\mathcal{P}_{\imath\jmath} (\vec{h}_{\imath\jmath},\vec{V}_{\imath\jmath})$ in the (D-2)-dimensional Euclidean space $\mathbb{R}_{\imath\jmath}^{D-2}$ in which $\mathcal{P}_{\imath\jmath} (\vec{h}_{\imath\jmath},\vec{V}_{\imath\jmath})$ lies, and $\vec{h}_{\imath\jmath}=(...,h_{\imath\jmath\jmath'},...)$ is the set of distances of the (D-3)-skeletons of $\mathcal{P}_{\imath\jmath} (\vec{h}_{\imath\jmath},\vec{V}_{\imath\jmath})$ from the
projection of the origin on $\mathbb{R}_{\imath\jmath}^{D-2}$. This operation can be continued so that arbitrary $d$-dimensional ($1<d<D$) skeletons of the original D-polytope can be obtained. Such a procedure can be illustrated as
\begin{equation}\label{dP}
\mathbf{L}_{\imath_{(D-d)}}^{d+1}\circ...\circ\mathbf{L}_{\imath_2}^{(D-1)}\circ\mathbf{L}_{\imath_1}^{D}: \quad \mathcal{P}(\vec{h},\vec{V})\Rightarrow\mathcal{P}_{\imath_1...\imath_{D-d}}(\vec{h}_{\imath_1...\imath_{D-d}},\vec{V}_{\imath_1...\imath_{D-d}}),
 \end{equation}
 where we defined
 \begin{equation}
\mathbf{L}_{\imath_{(D-d)}}^{d+1}:\quad \mathcal{P}_{\imath_1...\imath_{D-d-1}}(\vec{h}_{\imath_1...\imath_{D-d-1}}, \vec{V}_{\imath_1...\imath_{D-d-1}}) \Rightarrow \mathcal{P}_{\imath_1...\imath_{D-d}}(\vec{h}_{\imath_1...\imath_{D-d}},\vec{V}_{\imath_1...\imath_{D-d}}),
\end{equation} $\mathcal{P}_{\imath_1...\imath_{D-d}}(\vec{h}_{\imath_1...\imath_{D-d}},\vec{V}_{\imath_1...\imath_{D-d}})$ is the $(\imath_{D-d})_{\text{th}}$ $d$-skeleton of the $(d+1)$-polytope \\ $\mathcal{P}_{\imath_1...\imath_{D-d-1}}(\vec{h}_{\imath_1...\imath_{D-d-1}},\vec{V}_{\imath_1...\imath_{D-d-1}}) $, with $\vec{V}_{\imath_1...\imath_{D-d}}=(...,V^I_{\imath_1...\imath_{D-d+1}},...)$ being the set of the unit outward normal vectors of the $(d-1)$-skeletons of $\mathcal{P}_{\imath_1...\imath_{D-d}} (\vec{h}_{\imath_1...\imath_{D-d}},\vec{V}_{\imath_1...\imath_{D-d}})$ in the $d$-dimensional Euclidean space $\mathbb{R}_{\imath_1...\imath_{D-d}}^{d}$ in which $\mathcal{P}_{\imath_1...\imath_{D-d}} (\vec{h}_{\imath_1...\imath_{D-d}},\vec{V}_{\imath_1...\imath_{D-d}})$ lies, and $\vec{h}_{\imath_1...\imath_{D-d}}=(...,h_{\imath_1...\imath_{D-d+1}},...)$ is the set of distances of the $(d-1)$-skeletons of $\mathcal{P}_{\imath_1...\imath_{D-d}} (\vec{h}_{\imath_1...\imath_{D-d}},\vec{V}_{\imath_1...\imath_{D-d}})$ from the
projection of the origin on $\mathbb{R}_{\imath_1...\imath_{D-d}}^{d}$.
\subsection{Area of the d-skeleton}

To express the areas of arbitrary skeletons with $(h_\imath, V_\imath)$ of a D-polytope $\mathcal{P}(\vec{h},\vec{V})$, let us first consider the 3-dimensional skeletons $\mathcal{P}_{\imath_1...\imath_{D-3}}(\vec{h}_{\imath_1...\imath_{D-3}},\vec{V}_{\imath_1...\imath_{D-3}})$  in the 3-dimensional Euclidean space. $\mathcal{P}_{\imath_1...\imath_{D-3}}(\vec{h}_{\imath_1...\imath_{D-3}},\vec{V}_{\imath_1...\imath_{D-3}})$ can be given by a series of  algorithms as
 \begin{equation}\label{Ap}
\mathbf{L}_{\imath_{(D-3)}}^{4}\circ...\circ\mathbf{L}_{\imath_2}^{(D-1)}\circ\mathbf{L}_{\imath_1}^{D}: \quad \mathcal{P}(\vec{h},\vec{V})\Rightarrow\mathcal{P}_{\imath_1...\imath_{D-3}}(\vec{h}_{\imath_1...\imath_{D-3}},\vec{V}_{\imath_1...\imath_{D-3}}).
 \end{equation}
It has been shown in \cite{bianchi2011polyhedra} that the areas of the 2-faces of a 3-polyhedra can be given by the outward unit normal vectors and heights of these 2-faces. Then, by the results of $\vec{h}_{\imath_1...\imath_{D-3}}=\vec{h}_{\imath_1...\imath_{D-3}}(\vec{h},\vec{V})$ and $\vec{V}_{\imath_1...\imath_{D-3}}=\vec{V}_{\imath_1...\imath_{D-3}}(\vec{h},\vec{V})$ given by (\ref{Ap}), the area $\textrm{Ar}(\mathcal{P}_{\imath_1...\imath_{D-3}}(\vec{h}_{\imath_1...\imath_{D-3}} ,\vec{V}_{\imath_1...\imath_{D-3}}))$ of this 3-dimensional skeleton can be denoted by
  \begin{eqnarray}\label{3A}
A_{\imath_1...\imath_{D-3}}(\vec{h},\vec{V})&:=&\textrm{Ar}(\mathcal{P}_{\imath_1...\imath_{D-3}}(\vec{h}_{\imath_1...\imath_{D-3}} ,\vec{V}_{\imath_1...\imath_{D-3}}))=\frac{1}{3}\sum_{\imath_{D-2}} \textrm{Ar}(\mathcal{P}_{\imath_1...\imath_{D-2}})\cdot h_{\imath_1...\imath_{D-2}},
  \end{eqnarray}
  with $\textrm{Ar}(\mathcal{P}_{\imath_1...\imath_{D-2}})$ being the area of the $(\imath_{D-2})_{\text{th}}$ 2-faces of the 3-polyhedra $\mathcal{P}_{\imath_1...\imath_{D-3}}(\vec{h}_{\imath_1...\imath_{D-3}} ,\vec{V}_{\imath_1...\imath_{D-3}})$ as certain function of $(\vec{h}_{\imath_1...\imath_{D-3}} ,\vec{V}_{\imath_1...\imath_{D-3}})$ \cite{bianchi2011polyhedra}.
In turn, one can construct the expression of the area of an arbitrary $d$-face of the D-polytope $\mathcal{P}(\vec{h},\vec{V})$ with $\vec{h}$ and $\vec{V}$. The $d$-area of a $d$-polytope can be given by adding all the products of its $(d-1)$-faces' areas and heights and then dividing it by $d$, i.e.,
  \begin{equation}\label{Ah}
\text{Ar} (\mathcal{P}_{\imath_1...\imath_{D-d}}(\vec{h}_{\imath_1...\imath_{D-d}} ,\vec{V}_{\imath_1...\imath_{D-d}}))=\frac{1}{d}\sum_{\imath_{D-d+1}}\text{Ar}(\mathcal{P}_{\imath_1...\imath_{D-d+1}}(\vec{h}_{\imath_1...\imath_{D-d+1}} ,\vec{V}_{\imath_1...\imath_{D-d+1}}))\cdot h_{\imath_1...\imath_{D-d+1}},
  \end{equation}
  where $\text{Ar}(\mathcal{P}_{\imath_1...\imath_{D-d+1}}(\vec{h}_{\imath_1...\imath_{D-d+1}} ,\vec{V}_{\imath_1...\imath_{D-d+1}}))$ denotes the $(d-1)$-area of $(\imath_{D-d+1})_{\textrm{th}}$ $(d-1)$-face of the $d$-polytope $\mathcal{P}_{\imath_1...\imath_{D-d}}(\vec{h}_{\imath_1...\imath_{D-d}} ,\vec{V}_{\imath_1...\imath_{D-d}})$, and the sum is taken over all the $(d-1)$-faces of $\mathcal{P}_{\imath_1...\imath_{D-d}}(\vec{h}_{\imath_1...\imath_{D-d}} ,\vec{V}_{\imath_1...\imath_{D-d}})$. Thus by Eqs. (\ref{dP}), (\ref{3A}), (\ref{Ah}), it is easy to see that we can give the areas of arbitrary dimensional skeletons with $(\vec{V}, \vec{h})$ step by step as
  \begin{equation}\label{vf}
A_{\imath_1...\imath_{D-3}}(\vec{h},\vec{V}) \xrightarrow{\vec{h}_{\imath_1...\imath_{D-4}}} A_{\imath_1...\imath_{D-4}}(\vec{h},\vec{V})\xrightarrow{\vec{h}_{\imath_1...\imath_{D-5}}} ...\xrightarrow{\vec{h}_{\imath_1}} A_{\imath_1}(\vec{h},\vec{V}) \xrightarrow{\vec{h}}\textrm{Vol}(\mathcal{P}(\vec{h},\vec{V})),
  \end{equation}
  where $
A_{\imath_1...\imath_{D-d}}(\vec{h},\vec{V}):=\textrm{Ar}(\mathcal{P}_{\imath_1...\imath_{D-3}}(\vec{h}_{\imath_1...\imath_{D-3}} ,\vec{V}_{\imath_1...\imath_{D-d}}))$, and $\vec{h}_{\imath_1...\imath_{D-4}}=\vec{h}_{\imath_1...\imath_{D-4}}(\vec{h},\vec{V})$,..., $\vec{h}_{\imath_1}=\vec{h}_{\imath_1}(\vec{h},\vec{V})$ are given by the algorithm (\ref{dP}). Here we have denoted by $A_\imath$ the area of $\imath_{\textrm{th}}$ (D-1)-face of $\mathcal{P}(\vec{V}, \vec{h})$, which satisfies
  \begin{equation}\label{es}
A_\imath=A_\imath(\vec{V}, \vec{h}),\quad \forall \imath.
  \end{equation}
This system of equations can be solved for $h_\imath(\vec{A},\vec{V})$.  The existence of a solution with $h_\imath>0$, $\forall \imath$ is guaranteed by the Minkowski's theorem. However, the solution is not unique, because of the freedom of moving the origin around inside the polytope and thus changing the value of the heights without changing the shape of the polytope.

  Now the reconstruction of a D-polytope with its (D-1)- faces' areas and normal vectors is ready. For given $(\vec{A},\vec{V})$ satisfying Minkowski's theorem, we can choose a solution of $h_\imath(\vec{A},\vec{V})$ of the system of equations (\ref{es}), and then follow the Lasserres's reconstruction algorithm and equations (\ref{dP}), (\ref{3A}) and (\ref{Ah}) to obtain arbitrary d-dimensional skeletons $ \mathcal{P}_{\imath_1...\imath_{D-d}}(\vec{h}_{\imath_1...\imath_{D-d}},\vec{V}_{\imath_1...\imath_{D-d}})$ and their d-areas by $(\vec{V}, \vec{h}(\vec{A},\vec{V}))$ step by step. Especially, the D-volume $\textrm{Vol}(\mathcal{P}(\vec{h}(\vec{A},\vec{V}),\vec{V}))$  as a function of (D-1)-areas and normals has a number of interesting properties as follows.
 (i) The volume is by construction non-negative. For given (D-1)-areas, it vanishes if and only if the normals $(V_1^I,...,V_F^I)$ lie in a (D-1)-hyperplane including the special cases of $F <D+1$.
(ii) For fixed (D-1)-areas $\vec{A}$, the D-volume is a bounded function of the normals $(V_1^I,...,V_F^I)$. Let $V_{\textrm{max}}(\vec{A})$ be its maximum volume such that
      \begin{equation}
        V_{\textrm{max}}(\vec{A})\equiv\sup_{V^I_\imath} \{\textrm{Vol}(\mathcal{P}(\vec{h}(\vec{A},\vec{V}),\vec{V}))\}.
      \end{equation}
     Then $V_{\textrm{max}}(\vec{A})$ is smaller than the D-volume of the D-ball which has the same (D-1)-surface area as that of the D-polytope. Therefore we have the bound
      \begin{equation}
        0\leq V_{\textrm{max}}(\vec{A})<\frac{\pi^{\frac{D}{2}}(\frac{1}{2}\pi^{-\frac{D}{2}}(\sum_{\imath=1}^{F} A_\imath)\Gamma(\frac{D}{2}))^{\frac{D}{D-1}}}{\Gamma(\frac{D}{2}+1)},
      \end{equation}
      where $\Gamma$ is the Leonhard Euler's gamma function defined by $\Gamma(x)=\int_0^{+\infty}t^{x-1}e^{-t}dt, x>0$.
(iii) If one sets one of the (D-1)-areas to be zero so that the result is still a non-degenerate D-polytope, the D-volume function $\textrm{Vol}(\mathcal{P}(\vec{h}(\vec{A},\vec{V}),\vec{V}))$ given by (\ref{vf}) automatically measures the D-volume of the reduced D-polytope whose (D-1)-faces' number is $F-1$.

    It should be noted that a pair of $(\vec{A},\vec{V}^{IJ})$ with simplicity condition $V_\imath^{[IJ}V_\jmath^{KL]}=0$ determines uniquely the pair of $(\vec{A},\vec{V}^I)$ and hence the whole geometry of a D-polytope $\mathcal{P}(\vec{A},\vec{V}^{IJ})$  in $\mathbb{R}^{D+1}$. Note also that a point in $\mathfrak{P}^{s.}_{\vec{A}}$ which describes the shape of $\mathcal{P}(\vec{A},\vec{V}^{IJ})$ determines uniquely the equivalent class (up to $SO(D+1)$ rotation) of $\mathcal{P}(\vec{A},\vec{V}^{IJ})$ in $\mathbb{R}^{D+1}$. All of the discussion of the D-polytopes $\mathcal{P}(\vec{A},\vec{V})$ labelled by $(\vec{A},\vec{V})$ can be extended to the same polytopes $\mathcal{P}(\vec{A},\vec{V}^{IJ})$ labelled by $(\vec{A},\vec{V}^{IJ})$ equivalently.

    \subsection{Examples}
   To study concrete examples of D-polytopes we now consider the D-simplex and parallel D-polytope and discuss their volume expressions respectively.

   A D-dimensional simplex is a D-polytope with exactly $D+1$ vertices. Equivalently, it is the convex hull of a set of affinely independent $D+1$ points in $\mathbb{R}^D$. Denote by $(\ell^I_1,...,\ell^I_D)$ the oriented edges of a D-simplex $\mathcal{P}(\textrm{Sim},\vec{\ell}^I)$ with common beginning vertex, where the norm $|\ell_\imath^I|$ equals to the length of the $\imath_{\text{th}}$ edge. It is easy to see that this D-simplex is completely determined by the sets of vectors $(\ell^I_1,...,\ell^I_D)$ by definition. Then, the volume of $\mathcal{P}(\textrm{Sim},\vec{\ell}^I)$ is directly given by
   \begin{equation}
   \textrm{Vol}(\mathcal{P}(\textrm{Sim},\vec{\ell}^I))=\frac{1}{D!}|\bar{\epsilon}_{I_1I_2...I_D}\ell^{I_1}_1 \ell^{I_2}_2...\ell^{I_D}_D|,
   \end{equation}
   where $\bar{\epsilon}_{I_1I_2...I_D}$ is the Levi-Civita symbol in $\mathbb{R}^D$ in which $\mathcal{P}(\textrm{Sim},\vec{\ell}^I)$ locates.

   A parallel D-polytope $\mathcal{P}(\vec{A}, \vec{V}^I)_{p.}$ is a D-polytope whose (D-1)-faces' number is $2D$, and these (D-1)-faces are parallel in pairs such that $V_\imath^I=-V_{\imath+D}^I, \forall 1\leq\imath\leq D$ and each pair of parallel (D-1)-faces has same (D-1)-area, i.e., $A_\imath=A_{\imath+D}, \forall 1\leq\imath\leq D$. In fact, the parallel D-polytope $\mathcal{P}(\vec{A}, \vec{V}^I)_{p.}$ can be reconstructed by a set of its oriented edges $(\ell^I_1,...,\ell^I_D)$ with common beginning vertex and satisfying $V_I^\jmath\ell^I_\imath=0, \forall\jmath\neq\imath$. The relation between $(\vec{A}, \vec{V}^I)$ and $(\ell^I_1,...,\ell^I_D)$ can be illustrated as
   \begin{equation}\label{Al}
     A_\imath=A_{\imath+D}=\sqrt{\delta^{I_1J_1}\bar{\epsilon}_{I_1...I_D}\ell^{I_2}_1... \ell^{I_{\imath}}_{\imath-1}\ell^{I_{\imath+1}}_{\imath+1}...\ell^{I_D}_D\bar{\epsilon}_{J_1...J_D} \ell^{J_2}_1... \ell^{J_{\imath}}_{\imath-1}\ell^{J_{\imath+1}}_{\imath+1}...\ell^{J_D}_D},
   \end{equation}
   and
   \begin{equation}\label{Vl}
     V_{I_1}^\imath=- V_{I_1}^{\imath+D}=\frac{\epsilon_{I_1...I_D}\ell^{I_2}_1... \ell^{I_{\imath}}_{\imath-1}\ell^{I_{\imath+1}}_{\imath+1}...\ell^{I_D}_D} {\sqrt{\delta^{I_1J_1}\epsilon_{I_1...I_D}\ell^{I_2}_1... \ell^{I_{\imath}}_{\imath-1}\ell^{I_{\imath+1}}_{\imath+1}...\ell^{I_D}_D \epsilon_{J_1...J_D}\ell^{J_2}_1... \ell^{J_{\imath}}_{\imath-1}\ell^{J_{\imath+1}}_{\imath+1}...\ell^{J_D}_D}}.
   \end{equation}
There are three expressions of the volume of the parallel D-polytope $\mathcal{P}(\vec{A}, \vec{V}^I)_{p.}$. The first one involves only the set of oriented edges $(\ell^I_1,...,\ell^I_D)$ by
   \begin{equation}\label{Vp1}
   \textrm{Vol}(\mathcal{P}(\vec{A}, \vec{V}^I)_{p.})=|\bar{\epsilon}_{I_1I_2...I_D}\ell^{I_1}_1 \ell^{I_2}_2...\ell^{I_D}_D|=\ell_1...\ell_D|\bar{\epsilon}_{I_1I_2...I_D}\bar{\ell}^{I_1}_1 \bar{\ell}^{I_2}_2...\bar{\ell}^{I_D}_D|,
   \end{equation}
   where $\bar{\ell}_\imath^I:=\frac{\ell_\imath^I}{\ell_\imath}$, and $\ell_\imath:=\sqrt{\ell_\imath^I\ell_\imath^J\delta_{IJ}}$.
  The second one is constructed by (D-1)-areas and normals as
  \begin{equation}\label{Vp2}
   \textrm{Vol}(\mathcal{P}(\vec{A}, \vec{V}^I)_{p.})=\sqrt[D-1]{|\bar{\epsilon}_{I_1I_2...I_D}A_1...A_DV^{I_1}_1 V^{I_2}_2...V^{I_D}_D|}.
   \end{equation}
 The third expression is given by
   \begin{equation}\label{Vp3}
   \textrm{Vol}(\mathcal{P}(\vec{A}, \vec{V}^I)_{p.})=|A_\imath V_\imath^I\ell^J_\imath\delta_{IJ}|=|A_\imath\ell_\imath V_\imath^I\bar{\ell}^J_\imath\delta_{IJ}|, \quad\forall \imath.
   \end{equation}
   It is easy to see that the expressions \eqref{Vp1} and \eqref{Vp3} are equivalent by Eqs.\eqref{Al} and \eqref{Vl}.
   Also, the equivalence between (\ref{Vp1}) and (\ref{Vp2}) can be checked by substituting $\ell_\imath=\frac{\textrm{Vol}(\mathcal{P}(\vec{A}, \vec{V}^I)_{p.})}{A_\imath V_\imath^I\bar{\ell}^J_\imath\delta_{IJ}}$ into Eq. (\ref{Vp1}) and using
   \begin{equation}
     \prod_{\imath}V_\imath^{I}\bar{\ell}_{\imath,I}=\bar{\epsilon}_{I_1I_2...I_D} V_1^{I_1}...V_{D}^{I_D}\bar{\epsilon}_{J_1J_2...J_D}\bar{\ell}_1^{J_1}...\bar{\ell}_{D}^{J_D}.
   \end{equation}
Note that the volume of D-simplex $\mathcal{P}(\textrm{Sim},\vec{\ell}^I)$ relates to that of D-polytope $\mathcal{P}(\vec{A}, \vec{V}^I)_{p.}$ by
   \begin{equation}
     \textrm{Vol}(\mathcal{P}(\textrm{Sim},\vec{\ell}^I))=\frac{1}{D!}\textrm{Vol}(\mathcal{P}(\vec{A}, \vec{V}^I)_{p.})=\frac{((D-1)!)^{\frac{D}{D-1}}}{D!}\sqrt[D-1]{|\bar{\epsilon}_{I_1I_2...I_D}\tilde{A}_1...\tilde{A}_DV^{I_1}_1 V^{I_2}_2...V^{I_D}_D|},
   \end{equation}
 where $\tilde{A}_\imath=\frac{1}{(D-1)!}A_\imath$ is the (D-1)-area of the (D-1)-face of $\mathcal{P}(\textrm{Sim},\vec{\ell}^I)$ which is parallel to the edges $\ell^{I}_1,...,\ell^{I}_{\imath-1},\ell^{I}_{\imath+1},...,\ell^{I}_D$ of $\mathcal{P}(\vec{A}, \vec{V}^I)_{p.}$.
  \section{The meaning of quantum polytope}
  Based on the classical properties of D-polytope, in the rest of the paper we will discuss the relevance of D-polytope for all dimensional LQG. The relation comes from the following observations \cite{Gaoping2019coherentintertwiner}.
 First, simple coherent intertwiners are the building blocks of simple spin-network states in all dimensional LQG and contribute a non-orthogonal basis of the vertex kinematic Hilbert space satisfying the simplicity constraint.
Second, simple coherent intertwiners space is the quantum correspondence of the shape space of D-polytopes with fixed (D-1)-faces' areas discussed in previous sections.
Therefore the simple coherent intertwiners can be understood as the states of quantum polytope, and simple spin-network states as a collection of quantum polytopes associated with each vertex. In this section we will establish the notion of quantum polytopes and show that a simple coherent intertwiner is peaked at the geometry of a classical polytope.
%discuss the relevance of this fact for the relation between semiclassical states of loop quantum gravity.
  \subsection{The quantum D-polytope}
  The shape space $\mathfrak{P}^{\text{s}.}_{\vec{A}}$ of D-polytopes composed of the elements in the phase space $Q_{D-1}(A_\imath)$ of bi-vectors $A_\imath V_\imath^{IJ}$. To obtain the quantum space corresponding to $\mathfrak{P}^{\text{s}.}_{\vec{A}}$, let us first consider the quantization of $Q_{D-1}(A_\imath)$.
It is helpful to introduce the following geometric quantization example which is closely related to the case that we are considering \cite{mladenov1985geometric}\cite{ii1981geometric}. Consider a particle moving on an unit D-dimensional sphere with energy $\varepsilon:=\frac{\tilde{\Delta}}{2}:=\frac{L_{IJ}L^{IJ}}{4}$. The phase space of its angular momentum $L_{IJ}$ is given by the compact Kahler manifold $Q_{D-1}$ with Kahler form $\Omega_\varepsilon:=\sqrt{2\varepsilon}\Omega$. The geometric quantization of this model gives the following condition \cite{mladenov1985geometric}
\begin{equation}
\sqrt{2\varepsilon}=N+\frac{D-1}{2}, \quad N=0,1,2,...
\end{equation}
which implies the existence of the corresponding quantum Hilbert space $\mathcal{H}^N(Q_{D-1},\Omega_\varepsilon)$ labelled by quantum number $N$. The dimension of $\mathcal{H}^N(Q_{D-1},\Omega_\varepsilon)$ is given by
\begin{equation}
\textrm{dim}(\mathcal{H}^N(Q_{D-1},\Omega_\varepsilon)) =\frac{(2N+D-1)(D+N-2)!}{(D-1)!N!},
\end{equation}
which equals to the dimension of the space $\mathfrak{H}^{N,D+1}$ of homogeneous harmonic functions on $S^D$ with degree $N$. Thus $\mathfrak{H}^{N,D+1}$ is equivalent to $\mathcal{H}^N(Q_{D-1},\Omega_\varepsilon)$ as the representation space of the quantum algebra of the angular momentum operator $\hat{L}_{IJ}=-\textbf{i}X_{IJ}$, which is isomorphic to the Lie algebra $so(D+1)$ and where $X^{IJ}$ is a $so(D+1)$ valued bi-vector given by $X^{IJ}=2\delta^{[I}_K\delta^{J]}_L$ in the defining representation space of $SO(D+1)$. Note that the eigenvalues of $SO(D+1)$ Casimir operator $\Delta:=-1/2X_{IJ}X^{IJ}$ read $N(N+D-1)$.
Now we can identify $\sqrt{2}AV^{IJ}$ and $A^2$ with the angular momentum $L^{IJ}$ and Casimir operator $\Delta$ in above example respectively. Then we will get the quantum Hilbert space $\mathcal{H}^N(Q_{D-1},\Omega_{A^2/2})$ of $Q_{D-1}(A)$ with the condition
\begin{equation}
A=N+\frac{D-1}{2}, \quad N=0,1,2,...
\end{equation}
We can also choose the equivalent quantum Hilbert space $\mathfrak{H}^{D+1,N}$. The eigenvalues $N(N+D-1)$ of the Casimir operator $\Delta$ become the spectrum of the operator $\widehat{A^2}$ here. Also, the classical bi-vector $\sqrt{2}AV^{IJ}$ is promoted as operator $-\mathbf{i}X^{ IJ}$.

As mentioned in section 2, the shape space $\mathfrak{P}^{\text{s.}}_{\vec{A}}$ of D-polytopes in $\mathbb{R}^{D+1}$ is not a phase space but a constraint surface in the
phase space $\mathfrak{P}_{\vec{A}}$ given by Eq.\eqref{Pphase}.
Thanks to the Guillemin-Sternberg's theorem \cite{guillemin1982geometric}, the quantization commutes with the reduction with respected to the closure constraint \eqref{clou}. Hence we can first quantize the unconstrained phase space $\times_{\imath=1}^FQ_{D-1}(A_\imath)$ and then reduce it at quantum level by extracting the subspace of $\otimes_{\imath=1}^{F}\mathfrak{H}^{N_\imath,D+1}$, which is invariant under the global rotations of a D-polytope induced by the closure constraint. This gives precisely the $SO(D+1)$ gauge invariant intertwiner space $\mathcal{H}_{\vec{N}}:=\textrm{Inv}[\otimes_{\imath=1}^{F}\mathfrak{H}^{N_\imath,D+1}]$. The commutativity is summarized by the following diagram
$$\xymatrix{}
\xymatrix@C=3.5cm{
\times_{\imath=1}^FQ_{D-1}(A_\imath)
\ar[d]^{\textrm{Symplectic\ reduction}} \ar[r]^{\textrm{Geometric\ quantization}} &\otimes_{\imath=1}^F\mathfrak{H}^{D+1}_{N_\imath} \ar[d]^{\textrm{Quantum\ reduction} }\\
 \mathfrak{P}_{\vec{A}} \ar[r]^{\textrm{Geometric\ quantization}}& \mathcal{H}_{\vec{N}}}.$$
 The quantization also promote the classical variables $\sqrt{2}A_\imath V^{IJ}_\imath$ in $\mathfrak{P}_{\vec{A}}$ as operators $-\mathbf{i}X_\imath^{IJ}$ acting on elements in $\mathcal{H}_{\vec{N}}$. Then we can impose the simplicity constraint weakly in the quantum space $\mathcal{H}_{\vec{N}}$, by using the so-called coherent intertwiner states composed of the $SO(D+1)$ coherent states of Perelomov type \cite{Gaoping2019coherentintertwiner}. The resulting weak solutions of quantum simplicity constraint constitute the so-called simple coherent intertwiners space $\mathcal{H}^{s.c.}_{\vec{N}}$, which has the following two important properties \cite{Gaoping2019coherentintertwiner}. (i) The gauge-fixed formulation of its elements, which is called gauge-fixed simple coherent intertwiners, vanishes the expectation values of the quantum simplicity constraint with minimal uncertainty for given quantum numbers $\vec{N}$, while the expectation values of the quantum simplicity constraint for the gauge invariant simple coherent intertwiners remain some quantum interference terms that tend to zero in large $N$ limit.
(ii) A simple coherent intertwiner labelled by $(\vec{N},\vec{V}^{IJ})$ in $\mathcal{H}^{s.c.}_{\vec{N}}$ gives a semiclassical D-polytope, with its (D-1)-faces' areas $A_\imath\propto\sqrt{N_\imath(N_\imath+D-1)}$ and its shape  being identical to the classical polytope $\mathcal{P}(\vec{A},\vec{V}^{IJ})$ with minimal uncertainty. Also, each point in the classical phase space $\mathfrak{P}^{s.}_{\vec{A}}$ corresponds to a simple coherent intertwiner in $\mathcal{H}^{s.c.}_{\vec{N}}$. In this sense, there is a ``Classical-Quantum correspondence '' between $\mathfrak{P}^{s.}_{\vec{A}}$ and $\mathcal{H}^{s.c.}_{\vec{N}}$.

The quantization and the weak imposition of simplicity constraints can be summarized by the following diagram
$$\xymatrix{}
\xymatrix@C=3.5cm{
 \mathfrak{P}_{\vec{A}}\ar[d]^{\textrm{Impose}\ V_\imath^{[IJ}V_\jmath^{KL]}=0, } \ar[r]^{\textrm{Geometric}\ \textrm{Quantization}}& \mathcal{H}_{\vec{N}} \ar[d]^{\textrm{Impose}\ X_\imath^{[IJ}X_\jmath^{KL]}=0,} \\
 \mathfrak{P}^{s.}_{\vec{A}} \ar[r]^{\textrm{Classical-Quantum}}_{\textrm{Corespondence}}& \mathcal{H}^{s.c.}_{\vec{N}}.}$$
Notice that one can re-scale the (D-1)-areas $A$ with a constant $\mathrm{c}$ without changing the above result. In fact, up to a dimensionful constant, the generators $-\mathbf{i}X_\imath^{IJ}$ of $SO(D+1)$ acting on each simple representation space $\mathfrak{H}^{D+1}_{N_\imath}$ can be understood as the quantization of the vectors $A_{\imath}V_{\imath}^{IJ}$. In all dimensional LQG the dimensionful constant should be chosen as $8\sqrt{2}\pi\beta(l_p^{(D+1)})^{D-1}$, where $\beta$ is the Babero-Immirzi parameter and $l_p^{(D+1)}=\sqrt[D-1]{\frac{\kappa\hbar}{16\pi}}$ is the Planck length in $(1+D)$-dimensional space-time with $\kappa$ being the gravitational constant. So we have
\begin{equation}\label{qm}
A_{\imath}V_{\imath}^{IJ} \quad\longmapsto\quad\frac{1}{\sqrt{2}}\beta\kappa\hbar\mathbf{i}X^{IJ}_\imath =8\sqrt{2}\pi\beta(l_p^{(D+1)})^{D-1}\mathbf{i}X^{IJ}_\imath,
\end{equation}
and accordingly the closure constraint and simplicity constraint are promoted formally to operator equations,
\begin{equation}\label{GS1}
\sum_{\imath=1}^FA_\imath V^{IJ}_\imath=0 \longmapsto\sum_{\imath=1}^FX^{IJ}_\imath=0,
\end{equation}
\begin{equation}\label{GS2}
 V_\imath^{[IJ}V_\jmath^{KL]}=0\longmapsto X_\imath^{[IJ}X_\jmath^{KL]}=0.
\end{equation}
For the consideration of quantum D-polytopes, Eqs.\eqref{GS1} and \eqref{GS2} correspond to quantum Gaussian constraint and vertex simplicity constraint in (1+D)-dimensional LQG respectively. One can then proceed to associate operators to geometric observables through the quantization map (\ref{qm}). The area of a (D-1)-face of the quantum polytope is
\begin{equation}
  \hat{A}_\imath=\beta\kappa\hbar\sqrt{-\frac{1}{2}X_\imath^{IJ}X_{\imath,IJ}},
\end{equation}
with eigenvalues $A_\imath=16\pi\beta(l_p^{(D+1)})^{D-1}\sqrt{N_\imath(N_\imath+D-1)}$.
Similar to the case in standard (1+3)-dimensional LQG, the scalar product between the $SO(D+1)$ generators associated to two (D-1)-faces of a D-polytope measures the angle $\theta_{\imath\jmath}$ between them, i.e.,
\begin{equation}\label{angle}
  \hat{\theta}_{\imath\jmath}=\arccos\frac{\frac{1}{2}X_\imath^{IJ}X_\jmath^{IJ}} {\sqrt{(N_\imath(N_\imath+D-1))(N_\jmath(N_\jmath+D-1))}}.
\end{equation}
Notice that the angle operators do not commute among themselves. Therefore it is impossible to find a state for a quantum D-polytope with definite values of all the angles between its (D-1)-faces. Moreover, the adjacent relations of the (D-1)-faces are not prescribed a priori. Thus Eq.\eqref{angle} might not even be a true dihedral angle of the D-polytope.

   \subsection{Simple coherent intertwiner}
  The simple coherent intertwiners are defined as the $SO(D+1)$-invariant projection of a tensor product of states $|N_\imath, V_\imath^{IJ}\rangle\in \mathfrak{H}^{D+1,N_\imath}$, i.e.,
   \begin{equation}
   ||\vec{N},\vec{V}^{IJ}\rangle\equiv \int_{SO(D+1)}dg \bigotimes_{\imath=1}^{F}g|N_{\imath},V_\imath^{IJ}\rangle=\int_{SO(D+1)}dg g^{\otimes F}|\vec{N},\vec{V}^{IJ}\rangle,
   \end{equation}
   where $|\vec{N},\vec{V}^{IJ}\rangle\equiv \bigotimes_{\imath=1}^{F}|N_{\imath},V_\imath^{IJ}\rangle$ is the gauge-fixed simple coherent intertwiner such that the closure constraint and simplicity constraint are weakly satisfied. Here the states $|N_\imath, V_\imath^{IJ}\rangle$  are $SO(D+1)$ coherent states of Perelomov type satisfying the peakedness property \cite{long2020perelomov}
    \begin{equation}
   \langle N_\imath, V_\imath^{IJ}|X^{IJ}|N_\imath, V_\imath^{IJ}\rangle=2\mathbf{i}N_\imath V_\imath^{IJ},
    \end{equation}
and the relative uncertainty
    \begin{equation}\label{uncertainty}
    \frac{|\triangle(<X^{IJ}>)|}{|<X^{IJ}>|}=\frac{\sqrt{D-1}}{\sqrt{N}}.
    \end{equation}
    It is clear that Eq.\eqref{uncertainty} tends to zero for $N\rightarrow \infty$ \cite{long2020perelomov}. The expectation values of the geometric operators composed of the flux operators with respect to the simple coherent intertwiners $ ||\vec{N},\vec{V}^{IJ}\rangle$ have the nice property of semiclassicality. For example, the expectation values of the non-commuting angle operators $\hat{\theta}_{\imath\jmath}$ reproduce the classical angles between (D-1)-faces of the D-polytope in the large $N$ limit as
    \begin{equation}
     \frac{\langle\vec{N},\vec{V}^{IJ}||\cos\hat{\theta}_{\imath\jmath}||\vec{N},\vec{V}^{IJ}\rangle} {\langle\vec{N},\vec{V}^{IJ}||\vec{N},\vec{V}^{IJ}\rangle}\approx V_\imath ^{IJ}V_{\jmath,IJ},
    \end{equation}
 and the dispersions are small in comparison with the expectation values.

An useful property of gauge invariant coherent intertwiners is that they provide a over-complete basis of the gauge invariant intertwiner space. This can be seen as follow. It has been shown that the building block $|N,V\rangle$ of the coherent intertwiner provides a resolution of identity in the space $ \mathfrak{H}^{D+1,N_\imath}$ as
\begin{equation}
  \dim\left(\mathfrak{H}_{D+1}^N\right)\int_{Q_{D-1}}dV|N,V\rangle\langle N,V|=\mathbb{I}_{\mathfrak{H}_{D+1}^N}.
\end{equation}
 Thus it can be extended as a resolution of identity in the space $\otimes_{\imath=1}^F\mathfrak{H}^{D+1}_{N_\imath}$ as
\begin{equation}\label{id1}
\left(\prod_{\imath=1}^F\dim(\mathfrak{H}_{D+1}^{N_\imath})\right)\int_{\times_\imath Q^\imath_{D-1}}\prod_{\imath=1}^FdV_\imath|\vec{N},\vec{V}\rangle\langle \vec{N},\vec{V}|=\mathbb{I}_{\otimes_{\imath=1}^F\mathfrak{H}^{D+1}_{N_\imath}}.
\end{equation}
Such a resolution of identity also exists in the gauge invariant coherent intertwiner space. To simplify the realization of this resolution, we can suppose a parametrization of the phase space $\mathfrak{P}_{\vec{A}}$.
The phase space structure ensures that we can parametrize $\mathfrak{P}_{\vec{A}}$ via $2F(D-1)-D(D+1)$ real numbers $Z$. Hence the gauge invariant coherent intertwiner $||\vec{N},\vec{V}^{IJ}\rangle$ satisfying $\sum_{\imath=1}^FN_\imath V_\imath^{IJ}=0$ can be rewritten as $|\vec{N},Z\rangle$ with $Z$ representing the equivalent class (up to the globle $SO(D+1)$ rotation) of $\vec{V}^{IJ}$. Then, we can immediately introduce the resolution of identity in the gauge invariant coherent intertwiner space $\mathcal{H}_{\vec{N}}$ as
    \begin{equation}\label{HNID}
 \mathbb{I}_{\mathcal{H}_{\vec{N}}}=\int_{\mathfrak{P}_{\vec{A}}}d\mu(Z) |\vec{N}, {Z}\rangle\langle\vec{N},{Z}|,
    \end{equation}
    wherein $d\mu(Z)$ is the natural measure on $\mathfrak{P}_{\vec{A}}$ which is compatible with the symplectic structure.

   \section{General geometric operators}
    The simple coherent intertwiner space is regarded as the quantum space of D-polytopes, and the geometric properties of these quantum polytopes should be given by geometric operators defined in this space. Two kinds of general spatial geometric operators based on the basic holonomy and flux operators in all dimensional LQG have been defined in \cite{Gaoping2019geometricoperators}. In fact, the general geometric operators with respect to D-polytopes can also be defined by the coherent states of spatial geometry such as the gauge invariant coherent intertwiners.
The geometric operators of quantum polytopes involving coherent states can be defined based on following two facts: (i) A simple coherent intertwiner state is labelled by a point of the shape space of D-polytopes with fixed (D-1)-faces' areas, and the corresponding wave function is peaked at this point. (ii) A point in the shape space of D-polytopes gives full geometric information of the corresponding D-polytope.

 Let us consider the shape space $\mathfrak{P}^{s.}_{\!\vec{A}}$. Recall that the phase space $\mathfrak{P}_{\vec{A}}$ is parametrized via $2F(D-1)-D(D+1)$ real numbers $Z$. They induce the coordinates of the space $\mathfrak{P}_{\vec{A}}^{s.}$ of dimension $F(D-1)-\frac{D(D+1)}{2}$, where these parameters $Z$ have to satisfy $n=F(D-1)-\frac{D(D+1)}{2}$ independent equations $S_1=0,...,S_{n}=0$ corresponding to the simplicity constraint. We denote by $\bar{Z}$ the independent parameters of $Z$ that can parametrize the points in $\mathfrak{P}_{\vec{A}}^{s.}$. For an arbitrary point $p\in \mathfrak{P}_{\vec{A}}^{s.}$ with parameters $\bar{Z}|_{p}$, there is an equivalent class (up to global $SO(D+1)$ rotations) of the set $\vec{V}^{IJ}_{p}$ satisfying the closure and simplicity conditions. However, the gauge invariant simple coherent intertwiner $||\vec{N},\vec{V}^{IJ}\rangle$ is independent of the gauge choice of $\vec{V}^{IJ}$ and hence it can be equivalently denoted by $|\vec{N}, \bar{Z}\rangle$ with $\bar{Z}$ or the equivalent class of $\vec{V}^{IJ}$ representing the same point in $\mathfrak{P}_{\vec{A}}^{s.}$. Now, let us consider certain geometric quantity $G$, such as the length, area or volume and so on, of certain skeletons of the D-polytopes given by points of $\mathfrak{P}^{s.}_{\!\vec{A}}$.  The geometric quantity $G$ is a function $G(\vec{A},\bar{Z})$ on $\mathfrak{P}^{s.}_{\!\vec{A}}$, and its expression can be given by the reconstruction procedures introduced in section 3. However, the integral in the identity \eqref{HNID} takes over the space $\mathfrak{P}_{\vec{A}}$ which contains the points without any D-polytope meaning. To overcome the obstacle, one could extend the function $G(\vec{A},\bar{Z})$ to take value on whole space $\mathfrak{P}_{\vec{A}}$ as
    \begin{equation}\label{extension}
    G(\vec{A},\bar{Z})\mapsto G(\vec{A},Z),
    \end{equation}
    where $G(\vec{A},Z)$ is a proper distribution peaked at the subspace $\mathfrak{P}^{s.}_{\!\vec{A}}$ such that
    \begin{equation}
    G(\vec{A},Z)|_{Z=\bar{Z}}=G(\vec{A},\bar{Z}), \ \  G(\vec{A},Z)|_{Z\neq\bar{Z}}=0
    \end{equation}
     and
     \begin{equation}\label{ZbarZ}
     \int_{\mathfrak{P}_{\!\vec{A}}}d\mu(Z)G(\vec{A},Z)f(Z)=\int_{\mathfrak{P}^{s.}_{\!\vec{A}}}d\tilde{\mu}(\bar{Z})G(\vec{A},\bar{Z})f(\bar{Z}),
     \end{equation}
    with arbitrary functions $f(Z)$ on $\mathfrak{P}_{\vec{A}}$ and the measure $d\tilde{\mu}( \bar{Z})$ on $\mathfrak{P}_{\vec{A}}^{s.}$ induced by the natural measure $d\mu(Z)$ on $\mathfrak{P}_{\vec{A}}$.
    Then the geometric operator related to $G$ can be defined in the simple coherent intertwiner space $\mathcal{H}_{\!\vec{N}}^{s.c.}$ by
    \begin{equation}\label{go}
      \widehat{G}=\int_{\mathfrak{P}_{\vec{A}}} d{\mu}({Z}) G(\vec{A},{Z})|\vec{N},{Z}\rangle\langle\vec{N},{Z}|
    \end{equation}
   with $A_\imath=16\pi\beta(l_p^{(D+1)})^{D-1}\sqrt{N_\imath(N_\imath+D-1)}$. Note that $\mathcal{H}_{\!\vec{N}}^{s.c.}$ is a sub-Hilbert space of the gauge invariant coherent intertwiner space $\mathcal{H}_{\!\vec{N}}$. Thus the action of \eqref{go} on a state $|\phi\rangle\in \mathcal{H}_{\!\vec{N}}^{s.c.}$ involves the inner product $\langle \vec{N}, Z|\phi\rangle$ in $\mathcal{H}_{\!\vec{N}}$. Because of Eq.\eqref{ZbarZ}, the result of the action is still a state in $\mathcal{H}_{\!\vec{N}}^{s.c.}$ as
\begin{equation}
\hat{G} |\phi\rangle=\int_{\mathfrak{P}_{\vec{A}}} d{\mu}({Z}) G(\vec{A},{Z})|\vec{N},{Z}\rangle\langle\vec{N},{Z}|\phi\rangle=\int_{\mathfrak{P}^{\text{s.}}_{\vec{A}}} d\tilde{\mu}(\bar{Z}) G(\vec{A},\bar{Z})|\vec{N},\bar{Z}\rangle\langle\vec{N},\bar{Z}|\phi\rangle.
\end{equation}
By definition, the  geometric operators constructed by \eqref{go} have the desired semiclassical property. Note that in one of the two strategies proposed in Ref. \cite{Gaoping2019geometricoperators} the geometric operators are totally composed of flux and volume operators. They are also expected to have well semiclassical behaviours for simple coherent intertwiners. We will take the volume as an example to compare the semiclassical properties of its different operator versions.
    \subsection{D-volume operator of quantum D-polytopes}

    The volume of D-polytopes is a well-defined function on their shape space $\mathfrak{P}_{\vec{A}}^{s.}$. Let $V(\vec{A},\vec{V}^{IJ}(\bar{Z}))$ be the volume of a D-polytope with (D-1)-faces of areas $A_\imath=16\pi\beta(l_p^{(D+1)})^{D-1}\sqrt{N_\imath(N_\imath+D-1)}$ and unit normal bi-vectors $\vec{V}^{IJ}(\bar{Z})$ (up to $SO(D+1)$ rotations). Consider the space $\mathcal{H}_{\!\vec{N}}^{s.c.}$ of simple coherent intertwiners corresponding to the shape space $\mathfrak{P}_{\!\vec{A}}^{s.}$. The volume of quantum polytopes can be defined in terms of the gauge invariant coherent intertwiners  $|\vec{N},{Z}\rangle$ and of the classical volume as
\begin{equation}
  \hat{V}=\int_{\mathfrak{P}_{\!\vec{A}}} d{\mu}({Z}) V(\vec{A},Z)|\vec{N},{Z}\rangle\langle\vec{N},{Z}|,
\end{equation}
  where the distribution $V(\vec{A},Z)$ on $\mathfrak{P}_{\!\vec{A}}$ is the extension of volume function $V(\vec{A},\vec{V}^{IJ}(\bar{Z}))$ on $\mathfrak{P}_{\!\vec{A}}^{s.}$ following \eqref{extension}. This D-volume operator in $\mathcal{H}_{\!\vec{N}}^{s.c.}$ can be regarded as the extended version of the 3-volume operator in \cite{bianchi2011polyhedra}, which is defined in the $SU(2)$ coherent intertwiner space of standard (1+3)-dimensional LQG. Similar to the $SU(2)$ case, the D-volume operator in $\mathcal{H}_{\!\vec{N}}^{s.c.}$ has following interesting properties.
First, the operator $\hat{V}$ is positive semi-definite, since
  \begin{equation}
    \langle\phi|\hat{V}|\phi\rangle=\int_{\mathfrak{P}_{\!\vec{A}}} d{\mu}({Z}) V(\vec{A},Z)|\langle\phi|\vec{N},{Z}\rangle|^2=\int_{\mathfrak{P}_{\!\vec{A}}^{s.}} d\tilde{\mu}(\vec{A},\bar{Z}) V(\vec{A},\bar{Z})|\langle\phi|\vec{N},\bar{Z}\rangle|^2,
  \end{equation}
  for every normalized state $|\phi\rangle$  in $\mathcal{H}_{\vec{N}}^{s.c.}$. This is a straightforward consequence of the fact that the classical volume $ V(\vec{A},\bar{Z})$ is a positive function. Moreover, $\hat{V}$ vanishes for $F<D+1$.
Second, $\hat{V}$ is a bounded operator in $\mathcal{H}_{\vec{N}}^{s.c.}$. Its norm $||\hat{V}||=\sup_{\phi}\langle\phi|\hat{V}|\phi\rangle/\langle\phi|\phi\rangle$ is bounded by the maximum value of the classical D-volume of a D-polytope with fixed (D-1)-faces' areas, i.e.,
      \begin{equation}
      \frac{\langle\phi|\hat{V}|\phi\rangle}{\langle\phi|\phi\rangle} =\int d\tilde{\mu}(\bar{Z}) V(\vec{A},\bar{Z})|\langle\phi|\vec{N},\bar{Z}\rangle|^2\leq \sup_{\mathfrak{P}_{\!\vec{A}}^{s.}}\{V(\vec{A},\bar{Z})\}\equiv V_{\max(\vec{A})},
      \end{equation}
      where $V_{\max(\vec{A})}$ is bounded by the D-volume of the D-ball whose (D-1)-sphere surface's (D-1)-area is given by $\sum_{\imath=1}^FA_\imath$.
  Third, suppose that the operator $\hat{V}$ is defined in the Hilbert space $\mathcal{H}_{\vec{N}}^{s.c.}$ corresponding to the quantum D-polytopes with $F+1$ (D-1)-faces associated to quantum numbers $(N_1,...,N_{F+1})$. In the case of $N_{F+1}=0$ and $F>D$, it coincides with the volume operator in $\mathcal{H}_{\vec{N}}^{s.c.}$ corresponding to the quantum D-polytopes with $F$ (D-1)-faces associated to quantum numbers $(N_1,...,N_{F})$. This is a consequence of the fact that the classical volume of a D-polytope with $F+1$ (D-1)-faces coincides with the volume of a D-polytope with $F$ (D-1)-faces and the same normal vectors, if one of the areas of the faces in the former is sent to zero.

 Moreover, using the fact that for large $N$ limit two coherent intertwiners become orthogonal \cite{long2020perelomov},
one can show that the expectation value of $\hat{V}$ on a coherent intetwiner $|\vec{N},\bar{Z}\rangle$ reproduces at large $N$ limit the volume of the classical polytope with shape $(\vec{A},\bar{Z})$, i.e.,
\begin{equation}
  \langle\hat{V}\rangle\equiv\frac{\langle\vec{N},\bar{Z}|\hat{V}|\vec{N},\bar{Z}\rangle} {\langle\vec{N},\bar{Z}|\vec{N},\bar{Z}\rangle}\approx V(\vec{A},\bar{Z}).
\end{equation}
Hence, in the large $N$ limit, the largest expectation value of $\hat{V}$ is given by the volume of the largest D-polytope in $\mathfrak{P}_{\!\vec{A}}^{s.}$.
\subsection{The usual D-volume operator}
The usual strategy of constructing D-volume operator in all dimensional LQG is based on the action of the basic flux operator on spin network states. By regularizing the classical expression of the D-volume to adapt to the graph $\gamma$ of a spin network state, one can obtain some well-defined regularized operator in the kinematical Hilbert space of all dimensional LQG. The regularization depends on the dimension of spacetime. By removing the regulator, the D-volume operator $\hat{V}^{\textrm{B.T.}}(R)$ of region $R$ is defined as \cite{bodendorfer2013newiii}
\begin{equation}\label{volume}
\hat{V}^{\textrm{B.T.}}(R) =\int_Rd^Dp\hat{V}(p)_\gamma,
\end{equation}
with $ \hat{V}(p)_\gamma = (\hbar\kappa\beta)^{\frac{D}{D-1}}\sum_{v\in V(\gamma)}\delta^D(p,v)\hat{V}_{v,\gamma}$, where, for $D+1$ being even,
\begin{equation}\label{volume1}
  \hat{V}_{v,\gamma} \equiv |c_{\textrm{reg.}}\frac{\mathbf{i}^D}{D!}\sum_{e_1,...,e_D\in E(\gamma),e_1\cap...\cap e_D=v}s(e_1,...,e_D)\hat{q}_{e_1,...,e_D}|^{\frac{1}{D-1}}
\end{equation}
with $\hat{q}_{e_1,...,e_D}\equiv  \frac{1}{2}\epsilon_{IJI_1J_1I_2J_2...I_nJ_n}R_e^{IJ}R_{e_1}^{I_1K_1}R_{e'_1K_1}^{J_1}... R_{e_n}^{I_nK_n}R_{e'_nK_n}^{J_n}$, the set $\{e_1,...,e_D\}$ of the edges of $\gamma$ being relabelled as $\{e,e_1,e'_1,...,e_n,e'_n\}$, and $\epsilon_{IJI_1J_1I_2J_2...I_nJ_n}$ being the Levi-Civita symbol in the $(D+1)$-dimensional internal space, and for $D+1$ being odd,
\begin{eqnarray}\label{volume2}
  \hat{V}_{v,\gamma} &\equiv& ( \hat{V}_{v,\gamma} ^I  \hat{V}_{I\,v,\gamma} )^{\frac{1}{2D-2}},\\\nonumber
  \hat{V}_{v,\gamma}^I &\equiv&c_{\textrm{reg.}} \frac{\mathbf{i}^D}{D!}\sum_{e_1,...,e_D\in E(\gamma),e_1\cap...\cap e_D=v}s(e_1,...,e_D)\hat{q}^I_{e_1,...,e_D},
\end{eqnarray}
with $ \hat{q}^I_{e_1,...,e_D}\equiv \epsilon_{II_1J_1I_2J_2...I_nJ_n}R_{e_1}^{I_1K_1}R_{e'_1K_1}^{J_1}... R_{e_n}^{I_nK_n}R_{e'_nK_n}^{J_n}$ and the set $\{e_1,...,e_D\}$ being relabelled as $\{e_1,e'_1,...,e_n,e'_n\}$. Here $E(\gamma)$ and $V(\gamma)$ denote the collections of edges and vertices of graph $\gamma$ respectively, $s(e_1,...,e_D)$ is the orientation function of the tangent vectors of the edges $(e_1,...,e_D)$ at $v$, and $R_e^{IJ}$ is the right invariant vector fields on $SO(D+1)$.
 There is still an undetermined pre-factor $c_{\textrm{reg.}}$ or $c^2_{\textrm{reg.}}$ under the (D-1)$_{\textrm{th.}}$ or  (2D-2)$_{\textrm{th.}}$ root in the expression of the volume operators \cite{ma2010new}. In fact, a similar undetermined pre-factor appears also in the expression of 3-volume operator in the standard (1+3)-dimensional LQG. Certain ``triad tests'' were taken in the (1+3)-dimensional theory to fixe this pre-factor and a consistency result has been obtained \cite{giesel2006consistency}\cite{Yang_2019}. However, it is difficult to extend such ``triad test '' to all dimensional case \cite{Gaoping2019geometricoperators}, because of the ambiguity introduced by the anomalous quantum simplicity constraint. By requiring the semiclassical consistency based on the semiclassical D-polytopes, the pre-factor can be fixed case by case. This can be demonstrated  by following two samples.

In the first sample, we consider the semiclassical behaviours of the D-volume operator $\hat{V}^{\text{B.T.}}(R)$ based on D-parallel polytopes. In this case, the ``semiclassical'' states for spatial geometry  are based on the D-hypercubic graph $\gamma$ dual to the D-parallel polytopes. Such D-hypercubic graph is equipped with $2D$-valent vertices, and the tangent vectors of the edges linked to a same vertex $v$ span a D-dimensional vector space at $v$ so that they are parallel in pairs. The states are ``semiclassical'' in the sense that a coherent intertwiner is chosen on each vertex of the graph $\gamma$ underlying the states. More precisely, the ``semiclassical'' states are chosen such that, (i) each pairs of edges linked to same vertex $v$ with parallel tangent vectors at $v$ are labelled with same quantum numbers $N_\jmath=N_{\jmath+D}$, $\jmath=1,...,D$, and (ii) each vertex is labelled with a gauge invariant simple coherent intertwiner $||\vec{N},\vec{V}^{IJ}\rangle_{p.}$ peaked at a point of $\mathfrak{P}_{\!\vec{A}}^{s.}$, which corresponds to a parallel D-polytope with $A_\imath=16\pi\beta(l_p^{(D+1)})^{D-1}\sqrt{N_\imath(N_\imath+D-1)}$, $N_\jmath=N_{\jmath+D}$ and $V_\imath^{IJ}=-V_{\imath+D}^{IJ}$, $\jmath=1,...,D$. Let $R(v)$ be a sufficiently small open region dual to the vertex $v$. Then, according to the calculations in \cite{long2020perelomov}, the expectation value of the D-volume $\hat{V}_{R(v)}$ with respect to $||\vec{N},\vec{V}^{IJ}\rangle_{p.}$ reads at zeroth order of $\hbar$ as
   \begin{eqnarray}\label{expectationV1}
   \langle\hat{V}_{R(v)}\rangle&=&(\hbar\kappa\beta)^{\frac{D}{D-1}}(\frac{\sqrt{2}^D}{2D!}c_{\textrm{reg.}}\prod_{\jmath=1}^D N_\jmath|\epsilon_{IJI_1J_1I_2J_2...I_nJ_n}V_1^{IJ}V_{2}^{I_1K_1}V_{3\ K_1}^{J_1}... V_{D-1}^{I_nK_n}V_{D\ K_n}^{J_n}|)^{\frac{1}{D-1}}\nonumber\\
   &=&(\hbar\kappa\beta)^{\frac{D}{D-1}}(\frac{1}{D!}c_{\textrm{reg.}}\prod_{\jmath=1}^D N_\jmath |N^I\epsilon_{IJI_1J_1I_2J_2...I_nJ_n}V_1^{J}V_{2}^{I_1}V_{3}^{J_1}... V_{D-1}^{I_n}V_{D}^{J_n}|)^{\frac{1}{D-1}}
   \end{eqnarray}
   for $D=2n+1$ is odd, and
  \begin{eqnarray}\label{expectationV2}
   \langle\hat{V}_{R(v)}\rangle&=&(\hbar\kappa\beta)^{\frac{D}{D-1}}(\frac{\sqrt{2}^D}{D!}c_{\textrm{reg.}}\prod_{\jmath=1}^D N_\jmath |N^I\epsilon_{II_1J_1I_2J_2...I_nJ_n}V_{1}^{I_1K_1}V_{2\ K_1}^{J_1}... V_{D-1}^{I_nK_n}V_{D\ K_n}^{J_n}|)^{\frac{1}{D-1}}\nonumber\\
   &=&(\hbar\kappa\beta)^{\frac{D}{D-1}}(\frac{1}{D!}c_{\textrm{reg.}}\prod_{\jmath=1}^D N_\jmath |N^I\epsilon_{II_1J_1I_2J_2...I_nJ_n}V_{1}^{I_1}V_{2}^{J_1}... V_{D-1}^{I_n}V_{D}^{J_n}|)^{\frac{1}{D-1}}
   \end{eqnarray}
   for $D=2n$ is even, where $N^I$ is the unit vector satisfying $N^{[I}{V_\imath}^{JK]}=0$, $\forall \imath=1,...,2D$,  $V_1^{I_1},V_2^{I_2},...,V_D^{I_D}$ is the D independent normal vectors of (D-1)-faces of the parallel D-polytope that span the D-dimensional vector space orthogonal to $N^I$. Notice that the variables $(\vec{N},\vec{V}^{IJ})$ labelling the simple coherent intertwiner correspond to the variables $(\vec{A},\vec{V}^{IJ})$ defining a D-polytope around the vertex $v$. Although the volume of a convex D-polytope with $F$ faces is in general a rather complicated function of the (D-1)-areas and normals, its expression simplifies greatly for parallel D-polytopes as discussed in section 3. The volume of a parallel D-polytope is
   \begin{equation}\label{classicalV}
     V=\sqrt[D-1]{\prod_{\imath=1}^D A_\imath|\overline{\epsilon}_{I_1I_2...I_D}V_1^{I_1}V_2^{I_2}...V_D^{I_D}}|,
   \end{equation}
   where $\overline{\epsilon}_{I_1I_2...I_D}:=N^I\epsilon_{II_1I_2...I_D}$. Thus, in order to match the expectation value \eqref{expectationV1} and \eqref{expectationV2} of D-volume operator \eqref{volume} with respect to $||\vec{N},\vec{V}^{IJ}\rangle_{p.}$ with the classical volume \eqref{classicalV} of the D-polytope, the regularization factor in the D-volume operator \eqref{volume} should be given by $c_{\textrm{reg.}}=D!$ in the case of the D-hypercubic graphs.

In the second sample, we consider the semiclassical behaviour of D-volume operator $\hat{V}^{\text{B.T.}}(R)$ based on D-simplexes. In this case the simple coherent intertwiners $||\vec{N},\vec{V}^{IJ}\rangle_{s}$ at the (D+1)-valent vertices $v$ satisfy $\sum_{\imath=1}^{D+1}N_\imath V_\imath^{IJ}=0$ and $V_\imath^{IJ}=\sqrt{2}N^{[I}V_\imath^{J]}$, where the normal vectors $\{V_\imath^{I}\}$ span a D-dimensional vector space. Then, according to the calculation in \cite{long2020perelomov}, the expectation value of $\hat{V}_{R(v)}$ with respect to $||\vec{N},\vec{V}^{IJ}\rangle_{s}$ reads at zero order of $\hbar$ as,
   \begin{eqnarray}\label{simexpectV1}
   \langle\hat{V}_{R(v)}\rangle&=&(\hbar\kappa\beta)^{\frac{D}{D-1}}(\frac{D+1}{2D!\sqrt{2}^D}c_{\textrm{reg.}}\prod_{\jmath=1}^D N_\jmath|\epsilon_{IJI_1J_1I_2J_2...I_nJ_n}V_1^{IJ}V_{2}^{I_1K_1}V_{3\ K_1}^{J_1}... V_{D-1}^{I_nK_n}V_{D\ K_n}^{J_n}|)^{\frac{1}{D-1}}\nonumber\\
   &=&(\hbar\kappa\beta)^{\frac{D}{D-1}}(\frac{D+1}{D!2^{D}}c_{\textrm{reg.}}\prod_{\jmath=1}^D N_\jmath |\bar{\epsilon}_{II_1J_1I_2J_2...I_nJ_n}V_1^{I}V_{2}^{I_1}V_{3}^{J_1}... V_{D-1}^{I_n}V_{D}^{J_n}|)^{\frac{1}{D-1}}
   \end{eqnarray}
   for $D=2n+1$ is odd, and
  \begin{eqnarray}\label{simexpectV2}
   \langle\hat{V}_{R(v)}\rangle&=&(\hbar\kappa\beta)^{\frac{D}{D-1}}(\frac{D+1}{D!\sqrt{2}^D}c_{\textrm{reg.}}\prod_{\jmath=1}^D N_\jmath |N^I\epsilon_{II_1J_1I_2J_2...I_nJ_n}V_{1}^{I_1K_1}V_{2\ K_1}^{J_1}... V_{D-1}^{I_nK_n}V_{D\ K_n}^{J_n}|)^{\frac{1}{D-1}}\nonumber\\
   &=&(\hbar\kappa\beta)^{\frac{D}{D-1}}(\frac{D+1}{D!2^D}c_{\textrm{reg.}}\prod_{\jmath=1}^D N_\jmath |\bar{\epsilon}_{I_1J_1I_2J_2...I_nJ_n}V_{1}^{I_1}V_{2}^{J_1}... V_{D-1}^{I_n}V_{D}^{J_n}|)^{\frac{1}{D-1}}
   \end{eqnarray}
   for $D=2n$ is even. The semiclassical consistency requires that the expectation value \eqref{simexpectV1} and \eqref{simexpectV2} be equal to the volume
 \begin{equation}
     \textrm{Vol}(\mathcal{P}(\vec{A},\vec{V}^{IJ})_{s})=\frac{((D-1)!)^{\frac{D}{D-1}}}{D!} \sqrt[D-1]{|\bar{\epsilon}_{I_1I_2...I_D}A_1...A_DV^{I_1}_1 V^{I_2}_2...V^{I_D}_D|},
   \end{equation}
   of the classical D-simplex $\mathcal{P}(\vec{A},\vec{V}^{IJ})_{s}$, where $A_\imath=16\pi\beta(l_p^{(D+1)})^{D-1}\sqrt{N_\imath(N_\imath+D-1)}$ is the (D-1)-area of its $\imath_{\textrm{th.}}$ (D-1)-face. Then, the regularization factor in the D-volume operator \eqref{volume} should be fixed as
   \begin{equation}
     c_{\textrm{reg.}}=\frac{2^D((D-1)!)^D}{(D+1)(D!)^{D-2}}
   \end{equation}
   in the case of D-simplexes.
  \section{Summary and discussion}
In previous sections a number of properties of D-polytopes have been introduced. A D-polytope can be uniquely identified by the areas and the normals of its (D-1)-faces. Hence the shape space $\mathfrak{P}_{\!\vec{A}}^{\text{s.}}$ of D-polytopes was defined by Eq.\eqref{shapespace} through the areas and bi-vector normals of their (D-1)-faces. This shape space was then extended as the phase space $\mathfrak{P}_{\!\vec{A}}$ by neglecting the simplicity constraints, so that the quantization was achieved. We also introduced the Lasserre's reconstruction algorithm as expression \eqref{dP} and discussed how to derive explicitly the areas of $d$-skeleton of a D-polytope from the (D-1)-areas and normals through the reconstruction procedure as \eqref{vf}. Some general properties of the D-volumes of D-polytopes were given, and the volumes of D-simplex and parallel D-polytope were discussed in details.

The relevance of D-polytopes to the quantum theory was the main issue which we discussed. The geometric quantization of the phase space $\mathfrak{P}_{\!\vec{A}}$ was transferred to shape space $\mathfrak{P}_{\!\vec{A}}^{\text{s.}}$ by imposing weakly the quantum simplicity constraints. The result is the gauge invariant simple coherent intertwiner space whose elements can be interpreted as coherent states of D-polytopes. The knowledge of the simple coherent intertwiners corresponding to D-polytopes and the Lasserre's reconstruction algorithm were then used to define a new kind of spatial geometric operators by Eq.\eqref{go} including the areas of $d$-skeletons ($1\leq d\leq D$) of a quantum D-polytope. By construction these new operators have the correct semiclassical limit. We compared the new version of D-volume operator based on the D-polytope with the one usually constructed in all dimensional LQG. It was shown by two examples that the consistent semiclassical limit of the usual volume operator with respect to the semiclassical D-polytopes can be obtained by fixing its undetermined regularization factor case by case.

The relation between classical shape space of D-polytopes and gauge invariant simple coherent intertwiner space is remarkable. Similar to the studies for standard (1+3)-dimensional LQG \cite{rovelli2010geometry}\cite{freidel2010twisted}\cite{Bianchi:2009ky} , it is hopeful to extend this relation so that one could geometrically parametrize the phase space underlying all dimensional LQG. Also, the semiclassical consistency check for D-volume operator is expected to be extended to other spatial geometric operators. Note that, the general spatial geometric operators in \cite{Gaoping2019geometricoperators} were constructed by the building block operator, whose action on a quantum state depends on the actions of the basic flux and holonomy operators, while the spatial geometric operators \eqref{go} involve the coherent intertwiners as well as the classical geometry of D-polytopes. Thus, a prior, there is no guarantee that the two constructions of a same geometric quantity would give a same operator. However, if one restrict the actions of the different versions of a geometric operator on the semiclassical states based on the graph dual to a D-polytopes, it is possible to check their semiclassical consistency. The notion of quantum D-polytopes may shed a new light on further developments of all dimensional LQG.

\section*{Acknowledgements}

We benefited greatly from our numerous discussions with Shupeng Song, Cong Zhang and Xiangdong Zhang.  This work is supported by the National Natural Science Foundation of China (NSFC) with Grants No. 11875006 and No. 11961131013.

\bibliographystyle{unsrt}

\bibliography{ref}                        %ref为.bib文件名

\end{document}